\documentclass[final,5p,times,twocolumn]{elsarticle}

\usepackage[utf8]{inputenc}
\biboptions{sort&compress}
\usepackage{xfrac}
\usepackage{amsmath}
\usepackage{amssymb}
\usepackage{braket}
\usepackage{graphicx}
\usepackage{xcolor}

\usepackage{lineno}
\modulolinenumbers[5]

\usepackage{hyperref}
\hypersetup{colorlinks=true,
  citecolor=blue,linkcolor=blue}
\usepackage[capitalise]{cleveref}

\journal{Physics Letters B}

\newcommand{\eq}{\!=\!}

\begin{document}

\begin{frontmatter}

\title{Configuration crossing, shape evolution, and
odd-proton polarization in yttrium isotopes}
\author[nrcn,tau]{N.~Gavrielov}
\ead{noamgavrielov@gmail.com}

\address[nrcn]{Department of Physics,
  Nuclear Research Center Negev,
  Be'er Sheva 84190, Israel}
\address[tau]{School of Physics and Astronomy, Tel Aviv 
University, Tel Aviv, Israel}

\begin{abstract}
The odd-mass $^\text{91--101}$Y isotopes provide a testing
ground for how an unpaired proton modifies an abrupt
collective structural evolution. A configuration-mixing
Bose-Fermi description shows that the lowest negative- and
positive-parity states undergo a crossing of normal and
intruder configurations near neutron number $N \eq 60$, 
intertwined with an evolution from weak coupling of a 
quasiparticle to a near-spherical core toward strong 
coupling to a deformed core. To isolate the role of the odd 
proton, a differential charge-radius polarization 
observable is introduced as the difference between the 
isotope shifts of the odd-mass chain with those of the 
corresponding even-even cores. Its pronounced peak at $N 
\eq 58$ and sign reversal at $N \eq 60$ reveal a localized 
modification of the core evolution in the critical region.
Energy levels, wave-function content, electromagnetic
moments, two-neutron separation energies, and coherent-state
energy surfaces support this interpretation.
\end{abstract}

\begin{keyword}
Interacting boson-fermion model \sep
Configuration mixing \sep
Intertwined quantum phase transitions \sep
Odd-mass nuclei \sep
Nuclear shape coexistence \sep
Nuclear charge radii
\end{keyword}

\end{frontmatter}

Atomic nuclei can often disclose the organizing principles
associated with the number of nucleons ($A$), the proton
number ($Z$), and the neutron number ($N$), as well as how
the nucleon orbitals evolve. As one varies the number 
nucleons, such principles can change abruptly in certain 
regions of the nuclear chart, leading to identifications of 
shape evolution, coexistence and crossing --- often serving 
as manifestations of different types of quantum phase 
transitions (QPTs) \cite{Cejnar2010}. The identification of 
such abrupt changes in nuclei with even or odd proton and 
neutron numbers can often follow similar principles, for 
example through wave-function analyses and related 
observables. In other cases, however, the presence of an 
unpaired nucleon requires important distinctions.
For odd-mass nuclei a distinction from the even-even case 
could be the half-integer total angular momentum values 
assigned for the levels, where the ground state can change 
its values and parity along a chain of nuclei (isotopes or 
isotones), suggesting an intrinsic change in the structure 
caused by a QPT.

QPTs in even-even nuclei have been long studied 
\cite{Cejnar2010}, with two main types that 
are investigated: a shape evolution within a single shell 
model configuration \cite{Iachello2010QPT}, and 
coexisting shapes associated with different configurations 
that also cross \cite{Frank2006}.
In the $A \approx 100$ region, the abrupt crossing 
between weakly and strongly deformed configurations is 
driven by the isoscalar proton-neutron interaction 
\cite{Federman1979}. The enhanced attraction between 
protons and neutrons occupying spin-orbit partner 
orbitals, $\pi(n\ell_{\ell\pm1/2})$ and 
$\nu(n\ell_{\ell\mp1/2})$, lowers the energy of deformed
configurations and compensates for their larger 
single-particle and pairing energies \cite{Gavrielov2019, 
Gavrielov2022}. The increasing occupation of these orbitals 
also reorganizes the effective single-particle energies, a 
mechanism identified in large-scale shell-model 
calculations as type~II shell evolution \cite{Togashi2016, 
Otsuka2016a}.

In odd-mass nuclei, QPTs were mainly studied through the 
shape evolution of an adjacent even-even nucleus that
serves as a collective core \cite{Petrellis2011a, 
Iachello2011b}. Recently, crossings between configurations 
associated with different shapes were explicitly identified 
in the odd-mass niobium \cite{Gavrielov2022c, 
Gavrielov2023a} and zirconium \cite{Gavrielov2025} 
isotopes, where their intrinsic shapes were identified 
individually with weak mixing \cite{Leviatan2025}, leading 
to the notion of intertwined quantum phase transitions 
(IQPTs) in odd-mass nuclei.
More generally, the unpaired nucleon is coupled to the 
collective degrees of freedom of the remaining nucleons.
This coupling can induce shape polarization, modify the 
balance between competing configurations, and influence the
development of deformation along an isotopic chain 
\cite{Iachello2011b, Schunck2010, Quan2018a}. The
structural evolution of an odd-mass system can therefore
differ from that of the neighboring even-even nuclei,
providing a means to examine how a single nucleon affects a
collective quantum phase transition.

The odd-mass $^\text{91--101}$Y isotopes ($Z\eq39$)
provide a direct test of these mechanisms, since a single
proton is coupled to the strontium cores across the abrupt 
structural change near $N \eq 60$. 
Laser-spectroscopy measurements reveal 
an abrupt increase in charge radius, accompanied by changes 
in electromagnetic moments \cite{Cheal2007}. Decay and 
fast-timing studies constrain the predominantly 
single-particle structure of
$^{97}$Y \cite{Lhersonneau1986, Buscher1990}, while
prompt-fission $\gamma$-ray spectroscopy established
rotational bands in $^{99,101}$Y \cite{Luo2005}, whose 
strong collectivity was subsequently quantified by lifetime 
measurements \cite{Hagen2017}.

Previous theoretical studies addressed selected parts of
this structural evolution. The positive-parity states of
$^{97}$Y were described within a single-configuration IBFM
calculation \cite{Brant1988}, the rotational structures of
$^{99,101}$Y were analyzed using a triaxial
particle-rotor-plus-quasiparticle model \cite{Luo2005}, and
the ground-state shapes and one-quasiproton configurations
along the yttrium chain were investigated using 
self-consistent mean-field methods 
\cite{Rodriguez-Guzman2011}. Together, these studies 
established the evolution from predominantly 
single-particle structure to deformed rotational motion, 
but did not provide a chain-wide spectroscopic description 
with explicit mixing between normal and intruder 
configurations. 
The adjacent even-even $^\text{90--100}$Sr 
isotopes have recently been identified within the IBM-CM as 
undergoing an IQPT \cite{Gavrielov2026a}, with 
a crossing of the normal and intruder configurations 
between $^{96}$Sr and $^{98}$Sr accompanied by a shape 
evolution within the intruder configuration. This 
description of the strontium cores provides the basis for 
investigating how the configuration crossing and shape 
evolution develop in the presence of an odd proton, and how 
its coupling polarizes the core across the transition.

In the IBFM \cite{IachelloVanIsackerBook}, a boson core
representing the adjacent even-even nucleus
\cite{IachelloArimaBook} is coupled to a fermion. In the
IBFM-CM \cite{Gavrielov2022c,Gavrielov2023a}, the total
Hamiltonian has the matrix form
\begin{equation}\label{eq:ham-bf}
\hat H
= \left [
\begin{array}{cc}
\hat H^{\rm A}(\eta^{\rm A}) & \hat V(\omega)
\\
\hat V(\omega)  & \hat H^{\rm B}(\eta^{\rm B})
\end{array}
\right]~,
\end{equation}
where A denotes the normal configuration with $N_b$
bosons and B the intruder configuration with $N_b+2$
bosons \cite{Duval1981,Duval1982}, with respective
parameters $(\eta^{\rm A},\eta^{\rm B})$ and a mixing
term $\hat V(\omega)$. Each diagonal entry is composed of
boson, fermion and Bose-Fermi parts. The boson Hamiltonian 
is constructed from $s$ and $d$ bosons, representing 
valence nucleon pairs, and has a normal configuration A 
with $N_b = n_d + n_s$ bosons and an intruder configuration 
with $N_b + 2$ bosons, where $n_d$ and $n_s$ are the $d$- 
and $s$-boson numbers, respectively. The fermion 
Hamiltonian consists of effective single-particle energies. 
The Bose-Fermi Hamiltonian consists of monopole, quadrupole 
and exchange terms with strengths $A^{(i)}_0$, 
$\Gamma^{(i)}_0$ and $\Lambda^{(i)}_0$ ($i\eq{\rm A,B}$), 
scaled by occupation amplitudes $(u_j,v_j)$ of the single 
particle orbits $j$, following the microscopic 
interpretation of the IBFM \cite{IachelloVanIsackerBook}. 
The eigenstates are superpositions of basis states in the 
$[N_b]$ and $[N_b+2]$ spaces, 
\begin{multline}\label{eq:wf-bf}
\ket{\Psi;J} =
\sum_{\alpha,L,j}C^{(N,J)}_{\alpha,L,j}
\ket{\Psi_{\rm A};[N_b],\alpha,L,j;J} \\
+ \sum_{\alpha,L_,j}C^{(N+2,J)}_{\alpha,L,j}
\ket{\Psi_{\rm B};[N_b+2],\alpha,L,j;J}~.
\end{multline}
Here, $\alpha$ are the quantum numbers from the boson 
Hamiltonian, $L$ ($j$) the boson (fermion) angular 
momentum, and $J$ the total angular momentum.
The occupation probabilities of configuration 
($v^2_{(N_i,J)}$), orbit ($v^2_{(j;N_i,J)}$) and $d$-boson 
number ($v^2_{(n_d;N_i,J)}$) content are obtained
\begin{subequations}
\begin{align}
\label{eq:prob_ibfm_bos}
v^2_{(N_i,J)} & =
\sum_{n_d}v^2_{n_d;(N_i,J)} = \sum_{j}v^2_{j;(N_i,J)}~, 
\quad i = {\rm A,B}\\
\label{eq:prob_ibfm_spe}
v^2_{(j;N_i,J)} & = 
\sum_{\alpha,L}|C^{(N_i,J)}_{\alpha,L,j}|^2~,\\
\label{eq:prob_ibfm_nd}
v^2_{(n_d;N_i,J)} & = 
\sum_{\tau,n_\Delta,j,L}|C^{(N_i,J)}_{n_d, \tau, 
n_\Delta,L,j}|^2~.
\end{align}
\end{subequations}
The occupation probability $v^2_{(N_i,J)}$ discloses the 
content of normal and intruder configurations; 
$v^2_{(j;N_i,J)}$ the content of single-particle orbitals; 
and $v^2_{(n_d;N_i,J)}$ the dominant U(5) boson symmetry, 
equivalent to a spherical shape (phase) in the geometrical 
interpretation of the interacting boson model 
\cite{IachelloArimaBook}. The latter indicates the degree 
of deformation.
A single dominant $n_d$ component corresponds to a
spherical state, while a broad distribution over several
$n_d$ values indicates a deformed state, see 
\cite{Gavrielov2023a, Gavrielov2025} for details. 

Electromagnetic transition operators contain boson and 
fermion parts. The boson effective charges and $g$~factors 
are taken from the strontium cores \cite{Gavrielov2026a}. 
The fermion effective charge $e_f \eq -1.0849~e$ is fixed 
by the measured quadrupole moment of the $9/2^+_1$ state of 
$^{93}$Y, $-0.64(8)$~$e$b \cite{NDS.112.1163.2011}, and the 
free value $g_s\eq5.5857~\mu_N$ is quenched by a factor 
$q\eq0.874$ determined from the $1/2^-$ ground state 
magnetic moments of $^\text{91--97}$Y.

The $^\text{91--101}$Y isotopes are described by coupling
a proton to the respective even-even $^\text{90--100}$Sr
cores. The boson Hamiltonians, boson numbers and mixing
strengths are taken from the IBM-CM analysis of
\cite{Gavrielov2026a}, where the normal A~configuration
corresponds to 0p-2h and the intruder B~configuration to
2p-4h proton excitations with respect to the $Z \eq 40$
subshell closure, with boson numbers $(N_b, N_b+2)$
ranging from $(2,4)$ at neutron number 52 to $(7,9)$ at
62. Single quasiparticle energies $\epsilon_j$ and
occupation probabilities $v^2_j$ are obtained from a BCS
calculation with the single particle energies of
\cite{Barea2009} and a pairing gap
$\Delta_{\rm F} \eq 0.89$~MeV. The latter is extracted from 
the binding energies of the adjacent even-even strontium 
isotopes, assuming for simplicity the same values for 
both configurations, and consistent with the odd-mass 
zirconium pairing gap \cite{Gavrielov2025}. The resulting 
quasiparticle energies and occupation amplitudes, given in 
the caption of \cref{tab:parameters}, enter the quadrupole 
and exchange terms of the boson-fermion interaction 
\cite{Gavrielov2023a}.

\begin{table}[t]
\centering
\caption{\label{tab:parameters}
Strengths in MeV of the Bose-Fermi interactions for the
negative-parity and positive-parity states, obtained from
a fit assuming $A^{(i)}_0 \eq A_0$, $\Gamma^{(i)}_0 \eq 
\Gamma_0$ and $\Lambda^{(i)}_0 \eq \Lambda_0$, where 
$i \eq{\rm A,B}$. From a BCS calculation, $\epsilon_j \eq 
0.000$, 0.482, 0.883 and 0.526~MeV and $v^2_j \eq 0.392$, 
0.885, 0.934 and 0.107 for the $2p_{1/2}$, $2p_{3/2}$, 
$1f_{5/2}$ and $1g_{9/2}$ orbits, respectively, and the 
Fermi energy is $\lambda_{\rm F} \eq 1.56$~MeV.}
\begingroup
\setlength{\tabcolsep}{4.5pt}
\begin{tabular}{lcccccc}
\hline
Neutron number & 52 & 54 & 56 & 58 & 60 & 62 \\
\hline
\multicolumn{7}{c}{Negative parity} \\
\hline
$A_0$       & $0.50$ & $1.00$ & $1.00$ &
$1.00$ & $-1.00$ & $-0.07$ \\
$\Gamma_0$  & 0.30 & 0.30 & 0.30 & 0.70 & 0.70 &
0.70 \\
$\Lambda_0$ & 0.82 & 0.86 & 1.91 & 3.00 & 1.40 &
2.04 \\
\hline
\multicolumn{7}{c}{Positive parity} \\
\hline
$A_0$       & 0.00 & 0.00 & 0.00 & 0.00 & $-0.21$ &
$-0.27$ \\
$\Gamma_0$  & 0.30 & 0.30 & 0.30 & 0.70 & 0.03 &
0.03 \\
$\Lambda_0$ & 0.82 & 0.86 & 1.91 & 3.00 & 7.67 &
10.04 \\
\hline
\end{tabular}
\endgroup
\end{table}

\begin{figure*}[tb!]
\centering
\includegraphics[width=1\linewidth]
{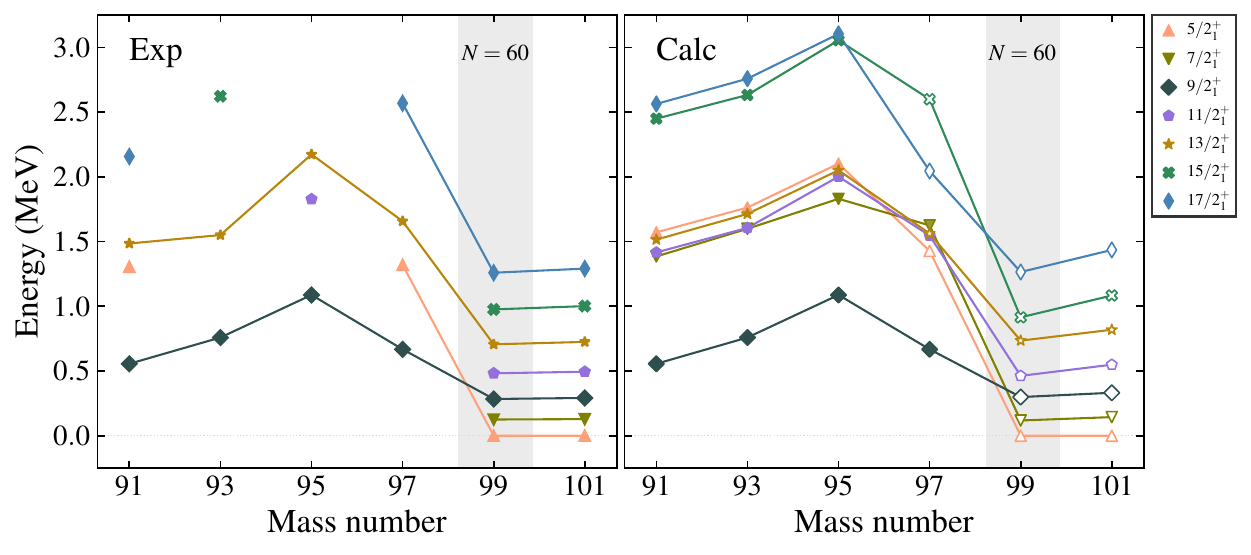}
\caption{Comparison between experimental (left panel) and
calculated (right panel) lowest positive-parity energy
levels along the odd-mass $^\text{91--101}$Y chain. In the 
calculation, the character of the symbols indicates the 
dominance of the normal A configuration (closed) or 
intruder B configuration (open). The shaded region marks 
neutron number 60. Data taken from \cite{ensdf}.
\label{fig:energies}}
\end{figure*}

The Bose-Fermi strengths obtained from the fit are listed
in \cref{tab:parameters}. Up to neutron number 58, the
quadrupole and exchange strengths are common to the two
parity sectors and vary smoothly, whereas the monopole term
acts mainly in the negative-parity multi-$j$ space to
readjust the relative quasiparticle multiplets. At neutron
number 60, $A_0$ becomes negative in both sectors, with a
larger magnitude for the negative-parity states, as the
low-lying structure reorganizes at the configuration
crossing. A pronounced parity dependence also develops at
neutron numbers 60--62: $\Gamma_0$ remains large and
$\Lambda_0$ moderate for the negative-parity states,
whereas the positive-parity sector requires a reduced
$\Gamma_0$ and an enhanced $\Lambda_0$. These bare
strengths cannot be compared directly because their
microscopic matrix elements contain different BCS
occupation factors. After this scaling is taken into
account, the effective positive-parity exchange coupling in
$^{99,101}$Y is comparable to that obtained in the niobium
isotopes \cite{Gavrielov2023a}.

We first consider the positive-parity sector.
\cref{fig:energies} shows the evolution of the lowest
positive-parity levels along the chain. For neutron
numbers 52--56, the lowest state is $9/2^+_1$, obtained
by coupling the $\pi 1g_{9/2}$ quasiparticle to the
$0^+_{1;\rm A}$ ground state of the spherical normal
configuration of the strontium core. The multiplet of states
$5/2^+,7/2^+,11/2^+,13/2^+$, obtained by coupling to the
$2^+_{1;\rm A}$ state, lies grouped above it around the
strontium-core $2^+_1$ energy, the pattern expected for weak
coupling to a spherical vibrator
\cite{IachelloVanIsackerBook}. From 58 the states
associated with the deformed intruder configuration drop
rapidly and at 60 the two configurations cross, as shown by 
the open symbols in the right calculation panel. The
lowest positive-parity state changes from
$9/2^+_1$ to $5/2^+_1$, reflecting the change from weak
coupling to a near-spherical core to strong coupling to a
deformed core. Outside a narrow region near the crossing, 
the states retain a high level of configuration purity. The 
calculated intruder probability of the lowest 
positive-parity state jumps from small values in 
$^\text{91--97}$Y to essentially unity in $^{99,101}$Y, in 
close analogy to the $0^+_1$ states of the strontium cores 
\cite{Gavrielov2026a}.

Beyond the crossing, the $5/2^+_1$ state becomes the
bandhead of the $K\eq5/2$ rotational ground-state bands
observed in $^{99,101}$Y. These bands have been assigned
to the $\pi5/2^+[422]$ Nilsson orbital and described by
particle-plus-rotor calculations with large, nearly axial
prolate deformation \cite{Luo2005}. The first five members
lie at 0, 125, 284, 482, and 706~keV in $^{99}$Y and at
0, 128, 292, 494, and 725~keV in $^{101}$Y. Fitting the
members from $5/2^+_1$ to $19/2^+_1$ with the linear
$J(J+1)$ expression
\begin{equation}\label{eq:rotor}
E(J) = E_0 + A_{\rm rot}J(J+1)~,
\end{equation}
gives $A_{\rm rot}\eq0.01751$ and $0.01798$~MeV for
$^{99}$Y and $^{101}$Y, respectively, with rms residuals
of 4--5~keV. These values are also comparable to the
rotational constant $B\eq0.018$~MeV extracted for the
well-developed $K\eq5/2$ band in $^{103}$Nb
\cite{Gavrielov2023a}, indicating similar moments of
inertia in the deformed Y and Nb structures. The calculated
Y sequences give $A_{\rm rot}\eq0.01670$ and
$0.01915$~MeV, differing from experiment by $4.6\%$ and
$6.5\%$, respectively. Their larger rms residuals of
25--31~keV indicate larger deviations from a linear
$J(J+1)$ dependence.
Altogether, the positive-parity structure evolves from weak
coupling of the $\pi1g_{9/2}$ single-quasiparticle to a
near-spherical core in $^\text{91--97}$Y to strong coupling
to a deformed core in $^{99,101}$Y.

\begin{figure}[tb!]
\centering
\includegraphics[width=0.96\linewidth]
{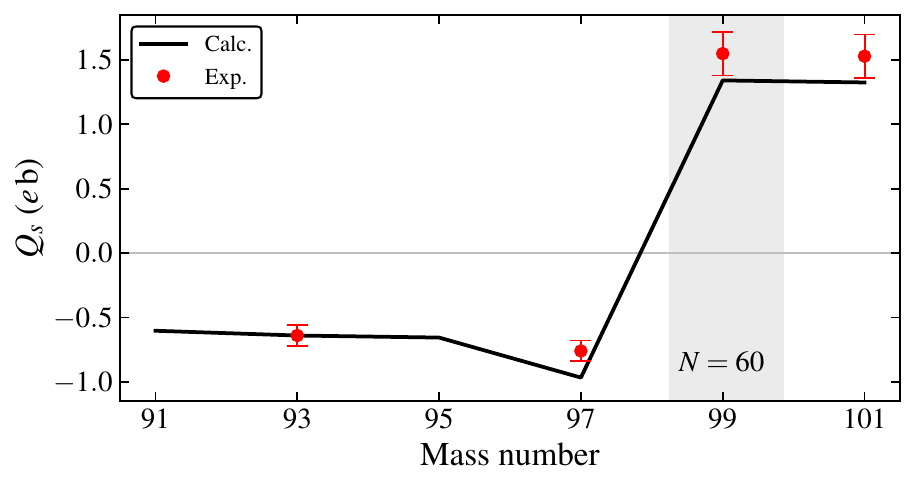}
\caption{Evolution of the spectroscopic quadrupole moments
$Q_s$ of the lowest positive-parity states along the
odd-mass Y chain. Symbols and lines denote experimental
data \cite{Cheal2007} and calculated results, respectively. 
The shaded region marks neutron number 60.
\label{fig:quadrupole_moments}}
\end{figure}

The negative-parity spectrum undergoes a similar structural 
evolution. For neutron numbers 52--58, the lowest states 
form the quasiparticle pattern associated with the 
$\pi(2p_{1/2},2p_{3/2},1f_{5/2})$ orbits weakly coupled to 
the near-spherical normal core. It consists of a $1/2^-$ 
ground state, low-lying $3/2^-_1$ and $5/2^-_1$ states, and 
$3/2^-_2$, $5/2^-_2$ states associated predominantly with 
coupling the $\pi2p_{1/2}$ quasiparticle to the $2^+_{1;\rm 
A}$ core excitation. At neutron number 60, the deformed 
intruder states become lowest also in this sector. The 
$d$-boson distribution of the lowest negative-parity state 
changes from a single dominant component with $\braket{\hat 
n_d}\eq0.01$ in $^{97}$Y to $\braket{\hat n_d}\eq6.7$ in 
$^{99}$Y, while its intruder probability increases from 
0.008 to 0.999. Experimentally, a low-lying group of 
negative-parity states of $^{99}$Y, $(5/2^-)$, $(3/2^-)$, 
$(7/2^-)$, and $(5/2^-)$ at 487, 536, 624, and 657~keV, is 
consistent with the calculated reorganization. The 
negative-parity sector therefore exhibits an abrupt 
configuration crossing accompanied by a shape evolution 
over the same neutron-number interval as the 
positive-parity sector.

The quadrupole moments of the lowest positive-parity states,
shown in \cref{fig:quadrupole_moments}, provide a direct 
spectroscopic measure for the evolution of deformation. The 
measured values are $-0.64(8)$ and $-0.76(8)~e{\rm b}$ in
$^{93,97}$Y \cite{Stone2005}, compared with calculated
values of $-0.64$ and $-0.97~e{\rm b}$. In $^{99}$Y,
$Q_s$ changes sign and increases to $+1.55(17)~e{\rm b}$ 
\cite{Cheal2007}, while the calculation gives $+1.34~e{\rm 
b}$ and predicts a similar $+1.32~e{\rm b}$ value in 
$^{101}$Y, with an experimental value of $+1.53(17)~e{\rm 
b}$. The sign change and large spectroscopic quadrupole 
moment therefore quantify the emergence of the strongly 
deformed prolate structure beyond the crossing.

The predominantly $\pi2p_{1/2}$ structure below the
crossing is also supported by the nearly constant measured
$1/2^-_1$ magnetic moments,
$-0.164(8)$, $-0.1390(9)$, $-0.16(3)$, and
$-0.12(1)~\mu_N$ in $^\text{91--97}$Y
\cite{Stone2005, Cheal2007}. The calculated
values are approximately $-0.146~\mu_N$ in
$^\text{91--95}$Y and $-0.13~\mu_N$ in $^{97}$Y, close
to the Schmidt value for a proton in the
$\pi2p_{1/2}$ orbit. 
A further test at high spin is provided by the $27/2^-_1$ 
isomer of $^{97}$Y, for which the calculation gives
$\mu\eq+5.63~\mu_N$ and $Q_s\eq-1.14$~$e$b, compared with
the measured $+5.64(4)~\mu_N$ and $-1.21(14)$~$e$b
\cite{Cheal2007}. This agreement provides a high-spin
benchmark for the electromagnetic operators.

\begin{figure*}[tbh!]
\centering
\includegraphics[width=1\linewidth]{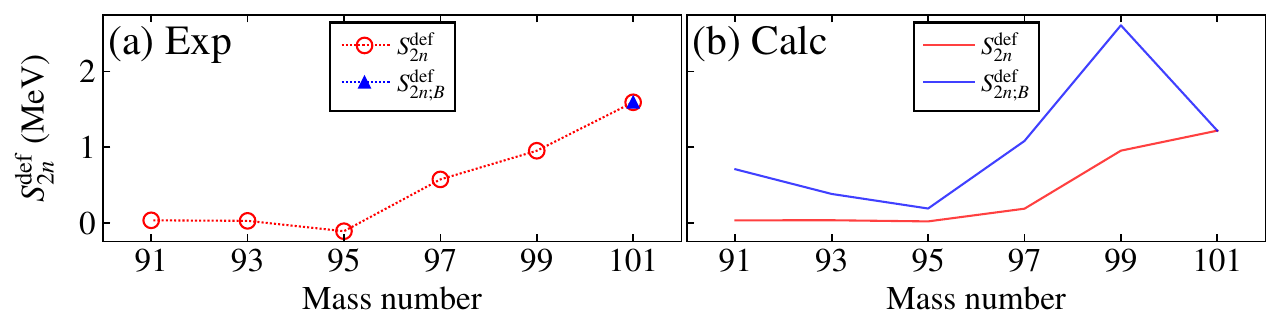}
\caption{Comparison of the deformed part of the
two-neutron separation energies, $S^{\rm def}_{2n}$,
between (a)~experiment \cite{Wang2021, Hager2007} and
(b)~calculation.
\label{fig:s2n}}
\end{figure*}

An observable that portrays both types of QPTs on the
ground states themselves is the two-neutron separation
energy, $S_{2n}\eq2M_n+M(N\!-\!2,Z)-M(N,Z)$, which can be
transcribed as \cite{Gavrielov2022,Gavrielov2023a}
\begin{equation}\label{eq:s2n}
S_{2n} = -\tilde A - \tilde B N_v \pm S^{\rm def}_{2n}
- \Delta_n~,
\end{equation}
In \cref{eq:s2n}, $N_v$ is half the number of valence 
particles in the boson core. $S^{\rm def}_{2n}$ is the 
contribution of the deformation obtained from the 
expectation value of the Hamiltonian in the ground state, 
the $+$ sign is for particles and the $-$ sign is for 
holes. The neutron subshell closure at $\nu 2d_{5/2}$ is 
reflected by $\Delta_n\eq0$~MeV for neutron numbers 52--58 
and 2~MeV for 58--62 as was taken in the even-even 
zirconium \cite{Gavrielov2019,Gavrielov2022} and odd-mass 
niobium \cite{Gavrielov2023a} isotopes. The linear 
parameters $\tilde A$ and $\tilde B$ are fixed in the 
smooth region below the transition, with values of 
$-16.266$ and 0.758~MeV, respectively. The value of $\tilde 
B$ is the same as for the even-even zirconium 
\cite{Gavrielov2019, Gavrielov2022} and odd-mass niobium 
\cite{Gavrielov2023a} isotopes. The value of $\tilde A$ is 
taken to fit $^{91}$Y. The constant values of $\Delta_n$ 
and $\tilde B$ for the strontium, yttrium, zirconium and 
niobium isotopes reflect the consistency of the 
\mbox{IBFM-CM} calculations for this region.
\cref{fig:s2n} shows the deformed part $S^{\rm def}_{2n}$, 
obtained by subtracting the linear part and $\Delta_n$ from 
the experimental \cite{Wang2021} and calculated 
$S_{2n}$. It is small for neutron numbers 52--56, where the 
ground states are spherical, and jumps at 58--60, where the 
onset of deformation occurs and the ground state becomes 
part of the deformed intruder configuration. 
The evolution of shapes and their crossing becomes more 
transparent as one calculates the two-neutron separation 
energy of the excited state associated with the lowest 
intruder positive-parity state. For that, one uses the mass 
of the excited state, $M_\text{exc}(N,Z) = M_\text{gs}(N,Z) 
+ E_\text{exc}(N,Z)$, where $M_\text{gs}(N,Z)$ is the 
ground state mass and $E_\text{exc}(N,Z)$ is the energy of 
the excited state (see \cite{Gavrielov2023a} for more 
details). This is depicted by the blue calculated line and 
experimental triangle (only in $A=101$ for lack of data) in 
\cref{fig:s2n}. The calculation suggests that the intruder 
states correspond to a quasiparticle weakly coupled to a
spherical core, which becomes more deformed at $A\eq97$
and undergoes an abrupt increase in deformation at
$A\eq\text{99--101}$. This behavior is very 
similar to that of promethium, europium and terbium 
isotopes, which were identified as undergoing a weak to 
strong coupling scenario of a single shape 
\cite{Iachello2011b, Petrellis2011a}.

\begin{figure}[tb!]
\centering
\includegraphics[width=1\linewidth]
{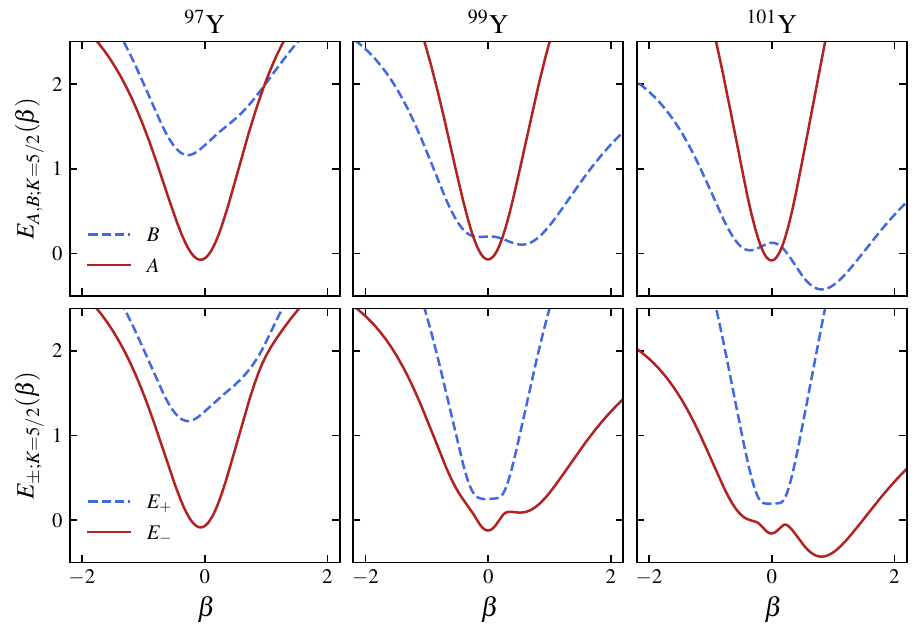}
\caption{Energy surfaces of the $K\eq5/2$ positive-parity
intrinsic states of $^{97,99,101}$Y as functions of the
deformation $\beta$, obtained as in \cite{Leviatan2025}.
Upper row shows the surfaces of the normal
A~configuration (solid) and intruder B~configuration
(dashed). Lower row shows the eigenpotentials obtained
after mixing.
\label{fig:geometry}}
\end{figure}

Recently, a geometric interpretation of the IBFM-CM was
developed and applied to the niobium isotopes
\cite{Leviatan2025}. Using this framework,
\cref{fig:geometry} shows the axial coherent-state energy
surfaces for the $K\eq5/2$ intrinsic states in
$^\text{97--101}$Y, spanning the region around the
critical point. Here, $\beta$ is the dimensionless
coherent-state deformation variable of the IBFM-CM. The
upper panels display the unmixed surfaces
$E_{{\rm A};K}(\beta)$ and $E_{{\rm B};K}(\beta)$ of the
normal and intruder configurations, respectively, while
the lower panels display the eigen-potentials
$E_{\pm;K}(\beta)$ obtained by diagonalizing the
corresponding coherent-state energy matrix
\cite{Leviatan2025}.

In $^{97}$Y, the lower eigen-potential has an almost
spherical, predominantly normal minimum. In $^{99}$Y, a
second, strongly prolate and almost purely intruder minimum
appears, while the near-spherical minimum remains the
global one. In $^{101}$Y, the prolate minimum becomes
deeper and replaces the near-spherical minimum as the
global minimum.
The appearance and increasing deformation of the prolate
minimum within the B~configuration portray the shape 
evolution. The change from a near-spherical to a prolate 
global minimum represents the mean-field manifestation of 
the configuration crossing. Its occurrence in $^{101}$Y, 
one isotope later than the crossing obtained in the exact 
finite-$N$ quantum calculation, reflects the smoothing of 
the abrupt phase-transitional behavior in the coherent-state
mean-field description, as found previously for the
even-even Zr isotopes.

Having established the configuration crossing and shape
evolution from both the quantum calculation and its
geometric interpretation, we now turn to charge radii,
which provide further information on the structure of the
selected low-lying states. The corresponding operator in
the IBFM-CM reads
\begin{equation}\label{eq:radius}
\hat T(r^2)
=
r_c^2+\alpha\hat N_b+\eta\hat n_d
+f_\pi\hat A^{(0)}_\pi~,
\end{equation}
where $r_c^2$ is the mean-square radius of the closed-shell
core, $\hat N_b$ and $\hat n_d$ are the boson- and
$d$-boson-number operators, respectively, and
\begin{equation}
\hat A^{(0)}_\pi 
=
-\sum_j 
\left[
a^\dagger_{\pi j}\times\tilde a_{\pi j}
\right]^{(0)}_0
=
-\sum_j\frac{\hat n_{\pi;j}}{\sqrt{2j+1}}~,
\end{equation}
is the scalar fermion operator with strength $f_\pi$.

For an odd-mass chain and its associated even-even core
chain, $\mathcal P=\pm$ denotes the selected parity
branch, and $J_N^{(\mathcal P)}$ is the total angular 
momentum of the lowest-energy state of parity $\mathcal P$ 
in the odd-mass isotope with neutron number $N$. The 
isotope shifts along the odd-mass branch and the even-even 
ground-state branch are defined, respectively, by
\begin{subequations}\label{eq:isotope_shifts}
\begin{align}
\Delta_{\mathcal P}\braket{\hat r^2}_{\rm odd}(N)
&\equiv
\braket{\hat r^2}_{{\rm odd};
J_{N+2}^{(\mathcal P)}}(N+2)
-
\braket{\hat r^2}_{{\rm odd};
J_N^{(\mathcal P)}}(N),
\label{eq:isotope_shift_odd}
\\
\Delta\braket{\hat r^2}_{\rm core}(N)
&\equiv
\braket{\hat r^2}_{{\rm core};0^+_1}(N+2)
-
\braket{\hat r^2}_{{\rm core};0^+_1}(N).
\label{eq:isotope_shift_core}
\end{align}
\end{subequations}
These differences cancel the closed-shell contribution
$r_c^2$ and isolate changes in the boson number,
$d$-boson content, and fermion contribution between
adjacent isotopes.
For the present calculation, the boson parameters were
determined for the strontium cores \cite{Gavrielov2026a} 
and are $\alpha \eq 0.235~{\rm fm}^2$, as in the even-even 
zirconium isotopes \cite{Gavrielov2019, Gavrielov2022}, and
$\eta\eq0.053~{\rm fm}^2$. The value of $f_\pi$ is set to
zero for lack of data. Nevertheless, this choice is also
revealing, as it removes the direct single-particle
contribution and leaves the odd proton to affect the radii
solely through its coupling to the boson core, thereby
isolating the core-polarization effect.

\begin{figure}[tb!]
\centering
\includegraphics[width=1\linewidth]
{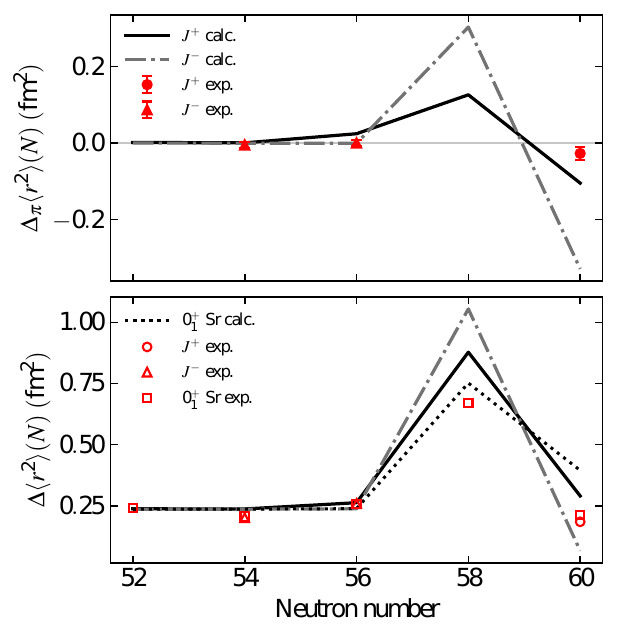}
\caption{The upper panel shows the differential
charge-radius polarization
$\Delta_\pi\braket{\hat r^2}_{\mathcal P}$ of
\cref{eq:delta_pi} for selected negative- and
positive-parity states in the odd-mass yttrium isotopes.
The negative-parity branch (gray dash-dotted line) follows
the $1/2^-_1$ states in $^{91\text{--}97}$Y, the
$5/2^-_1$ state in $^{99}$Y, and the $3/2^-_1$ state in
$^{101}$Y. The positive-parity branch (black line) follows
the $9/2^+_1$ states in $^{91\text{--}97}$Y and the
$5/2^+_1$ states in $^{99,101}$Y. The lower panel shows
the corresponding isotope shifts
$\Delta\braket{\hat r^2}$ of the odd-mass yttrium isotopes
and the even-even strontium cores. Symbols and lines denote
experimental data \cite{Cheal2007,Angeli2013} and
calculated results, respectively. Error bars were obtained
by propagating the quoted isotope-shift uncertainties in
quadrature.\label{fig:dpi}}
\end{figure}

For an odd-proton chain, the differential charge-radius
polarization is defined as the difference between
the isotope shift of the odd-mass system and that of its
even-even core,
\begin{equation}\label{eq:delta_pi}
\Delta_\pi\braket{\hat r^2}_{\mathcal P}(N)
\equiv
\Delta_{\mathcal P}\braket{\hat r^2}_{\rm odd}(N)
-
\Delta\braket{\hat r^2}_{\rm core}(N).
\end{equation}
In the present application, the odd-mass and core chains
are yttrium and strontium, respectively. The upper panel of
\cref{fig:dpi} shows
$\Delta_\pi\braket{\hat r^2}_{\mathcal P}(N)$ for their
selected negative- and positive-parity branches.
Since $f_\pi\eq0$, the differential polarization
originates entirely from differences between the evolution
of the boson-core components in the odd-mass system and
its even-even core. For either system $X={\rm odd,core}$,
the expectation value of the boson-number operator in a
state of angular momentum $I$ is
\begin{equation}\label{eq:Nb_expectation}
\braket{\hat N_b}_{X;I}(N)
=
N_A(N)+2v^2_{(N_B,I);X}(N),
\end{equation}
where $N_A$ and $N_B\eq N_A+2$ are the boson numbers of
the normal and intruder configurations, respectively, and
$v^2_{(N_B,I);X}(N)$ is the intruder-configuration 
probability defined in \cref{eq:prob_ibfm_bos} for an 
isotope with neutron number $N$. Using the same state 
selection as in \cref{eq:isotope_shifts}, we define
\begin{subequations}
\begin{align}
\Delta_{\mathcal P}v^2_{N_B;{\rm odd}}(N)
&\equiv
v^2_{(N_B,J_{N+2}^{(\mathcal P)});{\rm odd}}(N+2)
-
v^2_{(N_B,J_N^{(\mathcal P)});{\rm odd}}(N),
\\
\Delta v^2_{N_B;{\rm core}}(N)
&\equiv
v^2_{(N_B,0^+_1);{\rm core}}(N+2)
-
v^2_{(N_B,0^+_1);{\rm core}}(N),
\end{align}
\end{subequations}
and analogously for
$\Delta_{\mathcal P}\braket{\hat n_d}_{\rm odd}(N)$ and
$\Delta\braket{\hat n_d}_{\rm core}(N)$.
The normal-configuration boson number $N_A$ increases
identically between adjacent isotopes in the odd-mass and
core chains and therefore cancels in
\cref{eq:delta_pi}. Consequently,
\begin{equation}
\begin{split}
\Delta_\pi\braket{\hat r^2}_{\mathcal P}(N)
={}&
2\alpha
\left[
\Delta_{\mathcal P}v^2_{N_B;{\rm odd}}(N)
-
\Delta v^2_{N_B;{\rm core}}(N)
\right]
\\
&+
\eta
\left[
\Delta_{\mathcal P}\braket{\hat n_d}_{\rm odd}(N)
-
\Delta\braket{\hat n_d}_{\rm core}(N)
\right].
\end{split}
\label{eq:dpi_decomposition}
\end{equation}
The calculated values are essentially zero for
$N\eq52$, 54, and 56, where the yttrium and strontium
isotope shifts evolve nearly identically. At $N\eq58$,
$\Delta_{\mathcal P}v^2_{N_B;{\rm odd}}$ is 0.992 and
0.981 for the negative- and positive-parity branches,
respectively, compared with
$\Delta v^2_{N_B;{\rm core}}\eq0.885$. The corresponding
$2\alpha$ terms contribute 0.050 and
$0.045~{\rm fm}^2$, while the $\eta$ terms contribute
0.253 and $0.081~{\rm fm}^2$. Their sums give
$\Delta_\pi\braket{\hat r^2}_{\mathcal P}
\eq+0.303$ and $+0.126~{\rm fm}^2$, respectively.
At $N\eq60$, the selected yttrium states remain almost
purely intruder, whereas the intruder probability of the
strontium ground state continues to increase. The
$2\alpha$ terms are consequently $-0.049$ and
$-0.048~{\rm fm}^2$, and the $\eta$ terms are
$-0.279$ and $-0.056~{\rm fm}^2$, giving
$\Delta_\pi\braket{\hat r^2}_{\mathcal P}
\eq-0.329$ and $-0.105~{\rm fm}^2$ for the negative- and
positive-parity branches, respectively.
The lower panel of
\cref{fig:dpi} displays the underlying isotope shifts. The
yttrium and strontium values track one another below the
transition, separate at $N\eq58$, and exchange their
ordering at $N\eq60$. The available experimental values
follow the same ordering.

The peak is therefore not a smooth mass-dependent offset
between yttrium and strontium. It is localized in the
critical interval where the normal and intruder
configurations cross, and quantifies how the odd proton
modifies the evolution of the core configuration content
and deformation relative to the even-even strontium chain.

In conclusion, we have presented a unified IBFM-CM 
description of the odd-mass $^\text{91--101}$Y isotopes, 
obtained by coupling a proton to the corresponding 
$^\text{90--100}$Sr cores across the abrupt structural 
change near $N\eq60$. In both parity sectors, the lowest 
states undergo an abrupt crossing from the normal to the 
intruder configuration (Type~II QPT), intertwined with an 
evolution within the intruder configuration from weak 
coupling of a quasiparticle to a near-spherical core toward 
strong coupling to a deformed core (Type~I QPT). The level 
systematics, wave-function content, electromagnetic moments,
and two-neutron separation energies provide mutually
consistent signatures of this reorganization, while the
coherent-state energy surfaces display its geometric
counterpart in the competition between near-spherical and
prolate minima. These results establish the occurrence of
IQPTs in the odd-mass yttrium isotopes.
Together with the strontium, zirconium, and niobium chains 
\cite{Gavrielov2022, Gavrielov2022c, Gavrielov2025, 
Gavrielov2026a}, the present results extend the landscape 
of IQPTs 
in the $A\!\approx\!100$ region below the $Z\eq40$ subshell
closure.

The new differential charge-radius polarization introduced 
here adds to this picture by comparing the isotope shifts 
of an odd-mass chain directly with those of its even-even 
core. Its near-zero values below the transition, pronounced 
positive peak at $N\eq58$, and sign reversal at $N\eq60$ 
show that the odd-proton contribution is concentrated in 
the critical region where the normal and intruder 
configurations cross. The odd proton therefore 
modifies locally the evolution of the configuration content 
and deformation relative to the strontium core.
More broadly, the differential isotope-shift
construction provides a general route for using precision
charge-radius measurements to reveal how a single nucleon
modifies a collective core as its structure evolves.

\nolinenumbers
\bibliographystyle{elsarticle-num}
\bibliography{refs.bib}

\end{document}